\documentstyle[epsfig,seceqn]{elsart}
\journal{Nuclear Physics B}  
\def\etal{\hbox{\it et al.}}

\def\epj#1#2#3{    {\it Eur. Phys. J. }{\bf #1} (#2) #3}
\def\np#1#2#3{    {\it Nucl. Phys. }{\bf #1} (19#2) #3}
\def\npp#1#2#3{    {\it Nucl. Phys. }{\bf #1} 
Proceedings Supplement (19#2) #3}
\def\pl#1#2#3{    {\it Phys. Lett. }{\bf #1} (19#2) #3}
\def\jhep#1#2#3{    {\it JHEP }{\bf #1} (19#2) #3}
\def\pr#1#2#3{    {\it Phys. Rev. }{\bf #1} (19#2) #3}

\def\zp#1#2#3{    {\it Zeit. f\"ur Physik }{\bf #1} (19#2) #3}

\newcommand{\com}[1]{ \par }
\def\spa{\hspace*{-.65cm}}

\newcommand{\db}{\hspace{-0.2ex}\not\hspace{-0.7ex}D\hspace{0.1ex}}

\newcommand{\abs}[1]{\left| #1\right|} 

\def\evg{\, g_{eV}^\gamma}
\def\tvg{\, g_{\tau V}^\gamma}
\def\evz{\, g_{eV}^Z}
\def\eaz{\, g_{eA}^Z}
\def\tvz{\, g_{\tau V}^Z}
\def\taz{\, g_{\tau A}^Z}
\def\tz{\, g_{\tau T}^Z}
\def\tg{\, g_{\tau T}^\gamma}
\def\evg2{(g_{eV}^\gamma)^2}
\def\tvg2{(g_{\tau V}^\gamma)^2}
\def\evz2{(g_{eV}^Z)^2}
\def\eaz2{(g_{eA}^Z)^2}
\def\tvz2{(g_{\tau V}^Z)^2}
\def\taz2{(g_{\tau A}^Z)^2}
\def\tz2{(g_{\tau T}^Z)^2}
\def\tg2{(g_{\tau T}^\gamma)^2}

\def\re#1{{\mathrm Re}\left\{#1\right\} }
\def\im#1{{\mathrm Im}\left\{#1\right\} }
\def\eg{\epsilon_\gamma}
\def\ez{\epsilon_Z}
\def\eqs#1#2{eqs.~(\ref{#1}--\ref{#2})}
\newcommand{\beq}{\begin{equation}}
\newcommand{\eeq}{\end{equation}}
\newcommand{\bea}{\begin{eqnarray}}
\newcommand{\eea}{\end{eqnarray}}
\newcommand{\bes}{\begin{eqnarray*}}
\newcommand{\ees}{\end{eqnarray*}}   
\newcommand{\eq}[1]{eq.~(\ref{#1})}
\begin{document}
\noindent
FTUV$-$00$-$0218\hfill
IFIC/00$-$17
\begin{frontmatter}
\title{Model independent bounds on the tau lepton electromagnetic and weak
magnetic moments}
\author[Montevideo]{Gabriel A. Gonz\'alez-Sprinberg}, 
\author[Valencia]{Arcadi Santamaria} and 
\author[Valencia]{Jordi Vidal}
\address[Montevideo]{Instituto de F\'{\i}sica, Facultad de Ciencias,
Universidad de la Rep\'ublica, \\
Igu\'a 4225, 11400 Montevideo, Uruguay}
\address[Valencia]{Departament de F\'{\i}sica Te\`orica, IFIC, 
CSIC-Universitat de Val\`encia,\\ E-46100 Burjassot, Val\`encia, Spain} 
\begin{abstract}
Using LEP1, SLD and LEP2 data, for tau lepton production, and data from
$p\bar{p}$ colliders and LEP2, for W decays into tau leptons, we set model
independent limits on non-standard
electromagnetic and weak magnetic moments of the tau lepton.
The most general effective Lagrangian giving rise to tau moments is used
without further assumptions. Precise bounds ($2\sigma$) on the non-standard
model contributions to tau electromagnetic ($-0.007<a_\gamma< 0.005$),
tau Z-magnetic ($-0.0024 <a_Z< 0.0025$) and tau W-magnetic
($-0.003 < \kappa^W < 0.004$) dipole moments are set from the analysis.
\end{abstract}
\end{frontmatter}

\section{Introduction}
The present best bound on the tau lepton magnetic moment
($a_\gamma$) is indirect \cite{masso,pdg}. It comes
from the observation that in general extensions of the standard model (SM) 
it is very difficult to generate a magnetic moment for a lepton without 
originating a coupling of the Z boson to the lepton of the same order of 
magnitude. This
anomalous Z coupling ($a_Z$) is strongly bounded by LEP1, therefore,
by assuming that the magnetic moment of the lepton ($a_\gamma$) has 
the same size, 
one obtains a rather strong bound on it. 

While this argument is plausible, the complete amount of data coming
from tau-lepton production at LEP1, SLD and LEP2, and data on W decays into tau
leptons from LEP2 and $p\bar{p}$ colliders, allows for a more complete analysis of the
magnetic moment couplings of the tau to the photon, the Z and the W bosons. 

Following Ref.~\cite{masso,masso2}, in order to analyze tau magnetic moments, 
we will use an effective Lagrangian description\cite{buch}. Thus, 
in section \ref{sec:el} we 
describe the effective Lagrangian we use and fix the notation. LEP1 and SLD
are sensitive only
to the Z-magnetic moments, however LEP2 is sensitive to both Z- and photon-
magnetic moments. Far from the resonance statistics decreases dramatically
and the precision is not as good as the precision obtained at LEP1. 
However, since magnetic moment 
couplings are non-renormalizable couplings, their effects grow with energy
and, therefore, LEP2 limits can still be relevant. This is especially true
for electromagnetic couplings for which LEP1 does not give much information.
That is also the reason why 
experiments performed at lower energies do not provide, in general, 
stringent bounds (see for instance the bounds obtained from tau decays 
\cite{rizzo,lowbounds}). To obtain relevant bounds at lower energies
the suppression factor, $(E_{\mathrm low}/m_Z)^2$, has to be compensated 
by higher precision in the experiment.

Tau magnetic moment contributions to tau production
in LEP1-SLD and LEP2 are studied in section \ref{sec:tauprod}. In this respect
we can classify the observables in three classes: i) universality test
at LEP1, which are studied in \ref{sec:lep1}, ii) total production rates at
LEP2, considered  in \ref{sec:lep2}, and iii) transverse polarization asymmetries, 
studied in \ref{sec:pol}. 

Tau magnetic moments flip chirality and  in the standard model the only source
of
chirality flips are fermion masses. This means that any contribution of
magnetic moments (weak or electromagnetic) to total rates are either
suppressed by the fermion mass (relative to the electroweak scale) or need
two operator insertions and then,
they come as the square of the magnetic moments. In addition any new
physics, not only that related with magnetic moments, will appear in total
rates. Hence, in order to study magnetic moments it is interesting to
look for observables that are exactly zero when chirality is conserved.
Those observables will only be sensitive to fermion masses and to magnetic
moments. In addition they will only depend  linearly on magnetic moments. 
Some transverse tau polarization asymmetries \cite{nos} are observables of this
type and they have been
already measured at LEP1 \cite{acc} and SLD \cite{sld}. We will show 
in subsection
\ref{sec:pol} that they
already give now the best and cleanest bounds on tau magnetic moments. It
would be interesting to study these observables also at LEP2 to disentangle
weak an electromagnetic magnetic moments. 

Gauge invariance, after the
spontaneous symmetry breaking,  relates the
magnetic moments of the Z, the photon, and the W gauge bosons. Therefore 
one can also gain some insight on tau magnetic moments by studying
W decays into tau leptons. There are already rather good bounds on the 
universality of the leptonic decays of the W coming from LEP2, UA1, UA2, 
CDF and D0, those are studied in section \ref{sec:wdec}. In
section \ref{sec:limits} we combine all bounds and obtain the best
limits on the different magnetic moment couplings. Finally in section
\ref{sec:conc} we discuss the obtained results and compare with other
bounds found in the literature.

\section{The effective Lagrangian for tau magnetic moments\label{sec:el}}

The standard model gives a very good description of all physics
at energies available at present accelerators. Therefore, one expects
that any deviation from the standard model, at present energies, can be
parametrized by an effective Lagrangian built with the standard model
particle spectrum, having as zero order term just the standard model
Lagrangian, and containing higher dimension gauge invariant operators  
suppressed by the scale  of new physics $\Lambda$. The leading non-standard 
effects will come from the operators with the lowest dimension. Those are 
dimension six operators. There are only two operators of this type
contributing to the tau magnetic moments:
\begin{equation}
\label{eq:ob}
{\cal O}_B = \frac{g'}{2\Lambda^2} \overline{L_L} \varphi \sigma_{\mu\nu}
\tau_R B^{\mu\nu} ~,
\end{equation}
and
\begin{equation}
\label{eq:ow}
{\cal O}_W = \frac{g}{2\Lambda^2} \overline{L_L} \vec{\tau}\varphi
\sigma_{\mu\nu} \tau_R \vec{W}^{\mu\nu}  ~.
\end{equation}
Here $L_L=(\nu_L,\tau_L)$ is the tau leptonic doublet, 
$\varphi$  is the Higgs doublet, $B^{\nu\nu}$ and $\vec{W}^{\mu\nu}$ are the 
U(1)$_Y$ and SU(2)$_L$ field strength tensors, and $g'$ and $g$ are the gauge 
couplings. 

In principle, one could also write dimension six operators like
\begin{equation}
\label{eq:oderiv}
\frac{g'}{\Lambda^2} \overline{L_L} \sigma_{\mu\nu} \db L_L B^{\mu\nu} ~,
\end{equation}
with $ \db = \gamma_\mu D^\mu$. However, these operators
reduce to the operator \eq{eq:ob} after using the standard model 
equations of motion. In doing so, the couplings will be proportional 
to the tau-lepton Yukawa couplings. Note that operators in \eq{eq:ob} and
\eq{eq:ow} break chirality while the operator \eq{eq:oderiv} does not
break it. Therefore, by using this last form we would implicitly assume
that the only source of chirality breaking are Yukawa couplings and that
any chirality breaking, including magnetic moments, should be proportional
to fermion masses. In order to be more general we will assume the
forms in \eq{eq:ob} and \eq{eq:ow}, having in mind that we are introducing
in the standard model an additional source of chirality breaking in 
addition to fermion masses. As we will see in the next sections this will
become very important in looking for the right observables sensitive to
magnetic moments.

Thus, we write our effective Lagrangian as,
\begin{equation}
\label{eq:leff}
{\cal L}_{eff} = \alpha_B {\cal O}_B + \alpha_W {\cal O}_W + \mathrm{h.c.}~,
\label{eq:interaccio}
\end{equation}
where, for simplicity, we will take the couplings
$\alpha_B$ and $\alpha_W$ real. Note that complex couplings will break
$CP$ conservation and lead to electric dipole moments.

If these operators come from a renormalizable theory, in the perturbative
regime one expects, in general, that they arise only at one loop and therefore
their contribution will be further suppressed. However, this does not need to be 
the case, therefore  we leave the couplings $\alpha_B$ and
$\alpha_W$ 
as free parameters without any further assumption.

After spontaneous symmetry breaking, the Higgs gets a vacuum expectation
value $<\varphi^0>=u/\sqrt{2}$ with $u=1/\sqrt{\sqrt{2}G_F}=246$~GeV, and
the interactions (\ref{eq:interaccio}) can be written in terms of the gauge 
boson mass eigenstates, 
$A^\mu$ and $Z^\mu$, using that
\begin{eqnarray}
B^\mu    &=& -s_W Z^\mu + c_W A^\mu~, \nonumber\\
W_3^\mu  &=&  c_W Z^\mu + s_W A^\mu~,  \nonumber \\
W^{+\mu}  &=& \frac{1}{\sqrt{2}} \left(W_1^\mu-i W_2^\mu\right)~,
\end{eqnarray}
where, as usual, we define $c_W=\cos\theta_W$,  $s_W=\sin\theta_W$,
$\tan\theta_W = g'/g$ and $e=g s_W$. Thus, our Lagrangian, written in terms
of the mass eigenstates, is\footnote{Similar results are
found when using a non-linear realization of the gauge symmetry, 
as can be seen in Ref.\cite{masso2}}
\begin{eqnarray}
\nonumber
{\cal L}_{eff} &=& 
\epsilon_\gamma \frac{e}{2 m_Z}\overline{\tau} \sigma_{\mu\nu}\tau F^{\mu\nu}+
\epsilon_Z \frac{e}{2 m_Z s_W c_W}\overline{\tau} \sigma_{\mu\nu}
\tau
Z^{\mu\nu}\\ 
&+&
\left(\epsilon_W \frac{e}{2 m_Z s_W}  \overline{\nu_{\tau L}} 
\sigma_{\mu\nu}\tau_R W^{\mu\nu}_++\mathrm{h.c.}\right) ~,
\label{eq:leff_fin}
\end{eqnarray}
where $F_{\mu\nu}$ is the electromagnetic field strength tensor,
$Z_{\mu\nu}=\partial_\mu Z_\nu-\partial_\nu Z_\mu$ and
$W^{\mu\nu}_+=\partial_\mu W^+_\nu-\partial_\nu W^+_\mu$ are the
corresponding strength tensors  
for the $Z$ and $W$ gauge bosons. We have not written the non-abelian
couplings involving more than one gauge boson because they are not relevant
to our purposes.
We have normalized all couplings to $m_Z$, the natural mass-scale at the
energies we will consider, and we have defined the following
dimensionless couplings
\begin{eqnarray}
\epsilon_\gamma &=& (\alpha_B - \alpha_W) \frac{u
m_Z}{\sqrt{2}\Lambda^2}~,\\
\epsilon_Z &=& - (\alpha_W c_W^2 + \alpha_B s_W^2) \frac{u
m_Z}{\sqrt{2}\Lambda^2}~,\\
\epsilon_W &=& \alpha_W \frac{u m_Z}{\Lambda^2}=
-\sqrt{2}\left(\epsilon_Z+s_W^2 \epsilon_\gamma\right)~.
 \label{eq:epsilonw}
\end{eqnarray}
The Lagrangian (\ref{eq:leff_fin}) gives additional contributions to
the electromagnetic moment of the tau, which usually is expressed
in terms of the parameter $a_\gamma$. Similar parameters have been
introduced in the literature for the corresponding weak magnetic
moments for the $Z$-boson, $a_Z$ \cite{nos} and the $W$-boson, $\kappa^W$
\cite{rizzo}. They can be expressed in terms of our $\epsilon$'s
as follows,
\begin{eqnarray}
&&a_\gamma=\frac{2\, m_\tau}{m_Z}\, \eg ~,\\
&&a_Z=\frac{2\, m_\tau}{m_Z}\frac{1}{s_W\, c_W}\, \ez ~,\\
&&\kappa^W =  \sqrt{2} \frac{2\, m_\tau}{m_Z}\, \epsilon_W ~.
\label{eq:moments}
\end{eqnarray}
Notice that, in the effective Lagrangian approach, exactly the same couplings
that contribute to processes at high energies also
contribute to the magnetic moment form factors, $F^{\mathrm new}(q^2)$, at $q^2=0$. The
difference $F^{\mathrm new}(q^2)-F^{\mathrm new}(0)$ only comes from
higher dimension operators whose effect is suppressed by powers of 
$q^2/\Lambda^2$,  as
long as $q^2 \ll \Lambda$ as needed for the consistence of the effective
Lagrangian approach. 

In (\ref{eq:leff_fin}) we have all type of magnetic moment couplings except 
neutrino-neutrino couplings. This is due to the fact that we 
have not included right-handed neutrinos, $\nu_R$, in the particle spectrum.
If we include them we can add two additional operators that will give
magnetic moment couplings of the neutrinos to the Z, and additional
contributions to the $W^+$ magnetic moment couplings. However they
will also give rise to an electromagnetic moment for the neutrinos
which is extremely well bounded from a variety of sources, energy loss in 
red giants, supernova cooling, etc., so we are not going to consider them in
our calculation.

Since the effect of the operators in (\ref{eq:leff_fin})
is suppressed at low energies the most interesting bounds
will come from the highest precision experiments at the highest
available energies. Presently this means LEP1 and SLD
($Z$ decay rates and polarization asymmetries),
LEP2 (cross sections and $W$ decays rates), CDF and D0 ($W$ decay rates). 
Consequently in the following we will
study all those observables.

\section{$e^+ e^- \rightarrow \tau^+ \tau^-$  in presence of electromagnetic 
and weak magnetic moments\label{sec:tauprod}}
In this section we will consider $e^+ e^- \rightarrow \tau^+ \tau^-$
collision in a range of
energies from threshold to LEP2, therefore we will include both photon
 and Z-exchange with standard model vector and axial couplings to fermions, 
 plus additional
magnetic moment couplings given by \eq{eq:leff_fin}. The amplitude for
the process can be written as:
\begin{eqnarray}
{\cal M}\,&=&\, i\,e^2\,\sum_{k=\gamma,Z,...}
\bar{v}(p_2,s_2) \,\gamma_\mu(v_{e}^k-a_{e}^k \gamma_5)\,
u(p_1,s_1)\,\, {\cal
P}_k \nonumber \\&&
\bar{u}(p_3,s_3) \,\,[\gamma_\mu(v_{\tau }^k-a_{\tau }^k \gamma_5)\,
- \, g^k\,
 \frac{i}{2 m_\tau}\,\sigma^{\mu\nu}\,q_\nu]\,\,
v(p_4,s_4)  ~,
\label{eq:ampl_ee}
\end{eqnarray}
where the momenta $p_1$, $p_2$, $p_3$, $p_4$, correspond to
$e^-,e^+,\tau^-,\tau^+$
respectively, $q=p_1+p_2=p_3+p_4\,$, and ${\cal P}_k$ are the propagators of 
the different gauge bosons that contribute to the process:
\begin{eqnarray}
{\cal P}_\gamma = \frac{1}{q^2} &\hspace*{1cm};\hspace{1cm}&
{\cal P}_Z = \frac{1}{(2s_W c_W)^2}\,  \frac{1}{q^2 - m_Z^2+i\, 
m_Z \Gamma_Z } ~.
\label{eq:prop}
\end{eqnarray}
   
After squaring, summing and averaging over initial polarizations of 
electrons we obtain
\begin{eqnarray}
\overline{\sum} \abs{\cal M}^2 
&=& 
- \frac{1}{2} 
[
\re{{\cal V}^\mu {\cal V}^{*\nu}+ {\cal A}^\mu {\cal A}^{*\nu}} 
\,\,
(q^2 g_{\mu\nu}-q_\mu q_\nu+p_{i\mu}p_{i\nu}) 
+  \nonumber \\  & &
\im{{\cal V}^\mu {\cal A}^{*\nu}+{\cal A}^\mu {\cal V}^{*\nu}}
\,\, i\,
\epsilon_{\rho\sigma\mu\nu}\,q^\sigma \,p_i^\rho)]~,
\label{eq:ampl_ee2}
\end{eqnarray}
where  $\,p_i=p_2-p_1\,$, $\,p_f=p_4-p_3$, and
$\cal{V}$,
$\cal{A}$ carry all the coupling constants, final
spin and momentum dependences. After some algebra one finds:
\begin{eqnarray}
&&\hspace*{-1cm}{\cal V}^\mu = e\; 
\bar{u}(p_3,s_3)\sum_{k=\gamma,Z}v_e^i 
\left\{\gamma^{\mu}\left[(v^k_\tau-g^k)- a^k _\tau\gamma_5\right]-
\frac{1}{2m_\tau}  g^k \,p_f^\mu\right\}{\cal P}_k\, v(p_4,s_4) ~,\nonumber\\
&&\hspace*{-1cm} {\cal A}^\mu =e\;
\bar{u}(p_3,s_3) \sum_{k=\gamma,Z}a_e^i
\left\{\gamma^{\mu}\left[(a_\tau^k-g^k)-v^k_\tau\gamma_5\right]-
\frac{1}{2m_\tau}g^k\, p_f^\mu\right\}{\cal P}_k\, v(p_4,s_4). 
\label{eq:VA}
\end{eqnarray}
Here the couplings are:
\begin{eqnarray}
\displaystyle
&&\hspace*{-.75cm}v_{e}^\gamma = v_{\tau }^\gamma=\,- 1
\hspace*{3.35cm} ;\hspace*{1cm}  \displaystyle
a_{e}^\gamma= a_{\tau }^\gamma=\,0~,\\ \displaystyle
&&\hspace*{-.75cm}v_{e}^Z\equiv v_e = v_{\tau }^Z\equiv v_\tau=-\frac{1}{2}+2
s_W^2\hspace*{.25cm} ;\hspace*{1cm}\displaystyle
 a_{e}^Z\equiv a_e = a_{\tau}^Z\equiv a_\tau=-\frac{1}{2}~, \\ \displaystyle
&&\hspace*{-.75cm}g^\gamma = 
2\,\frac{m_\tau }{m_Z} \epsilon_\gamma
\hspace*{3.65cm};\hspace*{1cm}\displaystyle
 g^Z =  \frac{4 m_\tau}{m_Z} \epsilon_Z~.
\label{eq:g}
\end{eqnarray}

\subsection{Total rates and cross sections\label{sec:crossec}}
If we are not interested in final polarization we can sum over the
polarizations of the tau-leptons and obtain the angular distribution.
This is usually written in the following notation: 
\begin{eqnarray}
\hspace{-.7cm}\frac{d \sigma^0}{d\cos\theta} &=&
\sigma^0(s)\, \frac{1}{2}(1+\cos^2\theta)+
\sigma^m(s) \, \frac{1}{2}\sin^2\theta +
\sigma^{\rm FB}(s)\, \cos\theta~,
\label{eq:angdist}
\end{eqnarray}
where the coefficients $\sigma^i$ are given by:
\begin{eqnarray}
&&\hspace*{-.5cm} \sigma^0(s) =
\frac{\pi\alpha^2}{s}\beta
\left\{1+2\, v_\tau\, v_e\, \re{\chi}+(a_e^2+v_e^2)
(v_\tau^2+a_\tau^2\beta^2)\,
|\chi|^2\, \phantom{\frac{\frac{x}{x}}{X}}
{\frac{\phantom{}}{\phantom{}}}
\right.\nonumber\\
&&\hspace*{2.cm} \left.+ 4\, r_Z
\left[1+v_\tau\, v_e \re{\chi}\right]\eg
+4\, r_Z^2\, \epsilon_\gamma^2
\right.\nonumber \\
&&\hspace*{+2.cm} -8\, r_Z\, \left[
v_e\, \re{\chi}+v_\tau\, (v_e^2+a_e^2)\, |\chi|^2\right]\, 
\ez\label{eq:sigma0} \\
&&\hspace*{+2.cm}\left.+\left(4\, r_Z\right)^2
\,(v_e^2+a_e^2)\, |\chi|^2\,  
\epsilon_Z^2 - \left(4\, r_Z\right)^2\, v_e
\, \re{\chi}\, \ez\, \eg \phantom{\frac{1}{2}}\right\}~,\nonumber \\
&&\hspace*{-.5cm}\sigma^m(s)=
\frac{\pi\alpha^2}{s}\beta\left\{\frac{4\, m_\tau^2}{s}\left[
1+2 v_\tau\, v_e \re{\chi}+v_\tau^2(v_e^2+a_e^2)|\chi|^2\right]
\right.\nonumber\\
&&\hspace*{+2.5cm}+ 4\, r_Z
\left[1+v_\tau\, v_e \re{\chi}\right]\, \eg
+ \frac{s}{m_Z^2}\eg^2 \nonumber \\
&&\hspace*{2.5cm}\ -8\, r_Z
\left[v_\tau\, (v_e^2+a_e^2)\,
|\chi|^2+v_e\, \re{\chi}\right]\ez\label{eq:sigmam} \\
&&\hspace*{2.5cm}\left.+\frac{4\, s}{m_Z^2}(v_e^2+a_e^2)|\chi|^2\ez^2
-\frac{4\, s}{m_Z^2}\, v_\tau\, \re{\chi}\, \ez\, \eg
\right\}~,\nonumber\\
&&\hspace*{-.5cm}\sigma^{\rm FB}(s)=
\frac{\pi\alpha^2}{s}\beta^2\left\{
2\, a_\tau\, a_e\left[ \re{\chi}+2\, v_\tau\, v_e\, |\chi|^2\right]\right.\nonumber \\
&&\hspace*{2cm}\left. + 4\, r_Z\, a_\tau\, a_e\,\re{\chi}\, \eg
-16\, r_Z\, a_\tau\, a_e\, v_e\, |\chi|^2\, \ez\right\}~.
\label{eq:coef}
\end{eqnarray}
with
\begin{eqnarray}
&&r_Z=\frac{m_\tau}{m_Z},\quad\quad 
\beta =\beta(s)= \sqrt{1-\frac{4 m_\tau^2}{s^2}},\nonumber \\
&&\chi=\chi(s)=\frac{1}{(2s_Wc_W)^2}\, 
\frac{s}{(s-m_Z^2+i\,\Gamma_Z m_Z)},\nonumber
\end{eqnarray}
$s=q^2$, and $\theta$ being
the angle between the linear momentum of the $e^-$ and the $\tau^-$. 

In the limit of massless fermions ($\beta\rightarrow 1$) all linear terms
in the dipole moments disappear and the coefficients $\sigma^i$ simplify to:
\begin{eqnarray}
&&\hspace*{-.5cm} \sigma^0(s)|_{m_\tau\rightarrow 0} =
\frac{\pi\alpha^2}{s}
\left\{1+2\, v_\tau\, v_e\, \re{\chi}+(a_e^2+v_e^2)(v_\tau^2+a_\tau^2)\,
|\chi|^2\right\}~,\\
&&\hspace*{-.5cm}\sigma^m(s)|_{m_\tau\rightarrow 0}=
\frac{\pi\alpha^2}{s}\left\{ \frac{s}{m_Z^2}\, \left[ 
\eg^2 +\frac{4\, s}{m_Z^2}\, (v_e^2+a_e^2)\, |\chi|^2\, \ez^2\right.\right.
\nonumber\\ 
&&\hspace*{6.5cm}\left. \left.\phantom{\frac{s}{m_Z^2}}
-4\, v_\tau\, \re{\chi}\, \ez\,
\eg\right]\right\}~,\\
&&\hspace*{-.5cm}\sigma^{\rm FB}(s)|_{m_\tau\rightarrow 0}=
\frac{\pi\alpha^2}{s}\left\{\frac{\phantom{}}{\phantom{}}
2\, a_\tau\, a_e\left[ \re{\chi}+2\, v_\tau\, v_e\, |\chi|^2\right]\right\}~.
\label{eq:coef1}
\end{eqnarray}

It is worth noticing that $\sigma^m(s)$, the term in $\sin^2\theta$,
is proportional to the tau mass in the standard model and, therefore, it 
vanishes for  massless taus. 
However, in the presence of magnetic moment couplings
this term remains, even in the massless limit, and in fact it carries all the 
information about the magnetic moments $\ez$, $\eg$. 
This contribution peaks in the angular region where the standard model
contribution ($\frac{1}{2}(1+\cos^2\theta)\sigma^0|_{m_\tau\rightarrow 0}$) 
reaches its minimum.

By integrating out the angle in \eq{eq:angdist} we obtain the total cross section:
\begin{equation}
\hspace*{-.7cm}\sigma(e^+e^- \rightarrow \tau^+\tau^-) =
\frac{4}{3}\sigma^0(s)+\frac{2}{3}\sigma^m(s)~.
 \label{eq:crossec}
 \end{equation}

~From \eqs{eq:sigma0}{eq:coef1} we find that there are basically
three types of contributions:
i) the standard model contribution $\sigma^0|_{m_\tau\rightarrow 0}$, which
is the dominant one,
ii) a contribution which is proportional to the tau mass. This comes together
with an insertion of the magnetic moment operators (non-standard contribution 
to $\sigma^m$ and $\sigma^0$) or with an insertion of another fermion mass
(standard model contributions to $\sigma^m$) or both, two insertions of
magnetic moment operators and two mass insertions (non-standard
contributions to $\sigma^0$), and 
iii) a contribution free of masses but with two insertions of the magnetic
moment operators (in $\sigma^m$). 

This can be easily understood, since standard model couplings of gauge
bosons to fermions
conserve chirality, while mass terms and magnetic moment couplings change it,
therefore interference of magnetic moment contributions with standard
ones should be proportional to the fermion masses and only the square of
magnetic moments can be independent of fermion masses. In the limit of
zero tau mass only the contribution iii) is relevant, however there could
be some range of the parameters in which contribution ii) is higher
than iii). In fact for any finite value of the tau lepton mass it is obvious
that for large enough $\Lambda$ ii) will always dominate over iii). 
In order to be as general as possible we will include all three contributions.

\subsection{Bounds from $e^+ e^- \rightarrow \tau^+ \tau^-$ at LEP1 and SLD\label{sec:lep1}}
Bounds on the new couplings $\eg$ and $\ez$ can be obtained 
from LEP1-SLD universality tests by assuming that only the 
tau lepton has anomalous magnetic moments (muon and 
electron electromagnetic moments have been measured quite precisely
~\cite{pdg}). In order to compare with experimental data  it 
is convenient to define the universality ratio:

\begin{equation}
\hspace*{-.5cm}R_{\tau\mu}=\left.
\frac{\sigma(e^+e^-\rightarrow \tau^+\tau^-)}
                 {\sigma(e^+e^-\rightarrow
\mu^+\mu^-)}\right|_{s=m_Z^2}=
\frac{\Gamma_{\tau\bar{\tau}}}{\Gamma_{\mu\bar{\mu}}} =
\frac{R_\mu}{R_\tau} \equiv R_{SM} + R_1 +R_2~.
\label{eq:ratiocross}
\end{equation}
Here $R_\mu \equiv \Gamma_{\mathrm had}/\Gamma_{\mu\bar{\mu}}$ and
$R_\tau = \Gamma_{\mathrm had}/\Gamma_{\tau\bar{\tau}}$ are the quantities
measured directly \cite{lepsld}, on the other hand, $R_{SM}$ is the standard 
model contribution (including lepton-mass corrections), and
$R_1$ and $R_2$ are the linear and quadratic terms, respectively, in
the tensor couplings. 
In order to get bounds on the anomalous couplings, 
the theoretical  expression for $R_{\tau\mu}$ can 
be easily computed from \eqs{eq:angdist}{eq:coef1},
\begin{eqnarray}
&&R_{SM}=\sqrt{1-4\, r_Z^2}\, 
 \left[1+2\, r_Z^2\, \frac{v^2-2a^2}{v^2+a^2}\right]~, 
\nonumber\\
&&R_1=-12\, \sqrt{1-4\, r_Z^2}\, \,  r_Z\, 
\frac{v}{v^2+a^2}\, \ez~, \label{eq:ratio1}\\
&&R_2=2\, \sqrt{1-4\, r_Z^2}\, \left[1+8\, r_Z^2\right]
\,\frac{1}{v^2+a^2}\, \ez^2~.\nonumber
\end{eqnarray}
with $a\equiv a_e=a_\tau=a_\mu$,
and $v\equiv v_e=v_\tau=v_\mu$. Notice that in the ratio \eq{eq:ratiocross}
electroweak radiative corrections cancel to large extend 
and, therefore, we can use tree-level formulae. However, if needed,
the expressions in \eq{eq:ratio1} can be improved by using effective couplings 
\cite{bernabeu}.
Combining the 
very precise experimental LEP1 and SLD measurements \cite{lepsld},
\[
R_\mu = 20.786 \pm 0.033,\quad 
R_\tau = 20.764 \pm 0.045,\quad 
\]
we obtain
\begin{equation}
R_{\tau\mu}=1.0011 \pm 0.0027~.
\label{eq:ratiocross1}
\end{equation}

Comparing (\ref{eq:ratiocross1}) with (\ref{eq:ratiocross}) and 
(\ref{eq:ratio1})
one gets the condition for the  $\ez$ coupling to be
\begin{equation}
0.0007 \leq 7.967\, \epsilon_Z^2+0.037\, \epsilon_Z\leq 0.0061~.
\label{eq:epsilonz1}
\end{equation}
This equation leads to the following bounds on $\ez$: 
\begin{equation}\label{eq:epsilonz2}
-0.030\leq \epsilon_Z  \leq -0.012\quad \mathrm{or} \quad
0.007\leq \epsilon_Z \leq 0.025~,
\label{eq:ratiocross1limit}
\end{equation}
so that, at $1\sigma$, the zero value for $\ez$ is excluded. This is so because
the measured value of $R_{\tau\mu}=1.0011$, given in \eq{eq:ratiocross1},
excludes the SM (at $1\sigma$) due to the fact that the SM 
mass correction ($r_Z$) to $R_{\tau\mu}=1$ is negative while the measured 
value is larger than one. At $2\sigma$ the effect disappears.

It must be noticed that, in \eq{eq:epsilonz1}, the linear term in
$\epsilon_Z$ is strongly suppressed by the mass insertion
$r_Z$ --necessary to get a chirality
even contribution to the observable-- and also by the vector coupling $v$
($\frac{1}{4}$ effect), so that these bounds on $\epsilon_Z$ come
almost entirely from the quadratic term in the coupling.
Note also that, as expected from 
\eq{eq:ratio1}, no bound is found on the $\gamma$-coupling
$\epsilon_\gamma$, due to the $Z$ dominance at the $Z$-peak. 

\subsection{Bounds from 
$e^+ e^- \rightarrow \tau^+ \tau^-$ at {LEP2}\label{sec:lep2}}
The  situation is quite different at LEP2, where contributions coming
from the photon-exchange are dominant over those coming from
the $Z$-exchange. This is easily seen from the expression of the
$e^+e^-\longrightarrow \tau^+\tau^-$ cross  section given  in  
\eq{eq:sigma0} to \eq{eq:coef}.

Present limits from $e^+e^-\rightarrow \tau^+\tau^-$ cross section
are much milder at LEP2 \cite{all}. A combination of the LEP2 data on this
cross section, for a value
of $s'$  (the invariant mass of the pair of tau leptons) so that
$\sqrt{s'/s}>0.85$, is listed in table \ref{tabla1}. This 
combination of data has been only made for  the 183 GeV and 189 GeV data-sets as they have the highest
luminosities and center-of-mass energies. For comparison we also present
the standard model prediction for the cross section. In both, experimental
results and standard model predictions, initial-final state radiation photon
interference is subtracted.

\begin{table}
\caption{Combined experimental data for the $\tau^+\, \tau^-$ cross section 
from ALEPH, DELPHI, L3 and OPAL at LEP2 energies. $s'/s$ is cut in the invariant mass of the tau pair}
\begin{tabular}{|c||c|c|c|c|}
\hline
$\sqrt{s}$~(GeV)&$\sigma_{\tau\bar{\tau}}^{\rm
SM}$(pb)&$\sigma_{\tau\bar{\tau}}$(pb)&$\sqrt{s'/s}>$&Collaboration\\ \hline
\hline
182.7&3.45&$3.43\pm0.18$&0.85&LEP Electroweak\\ \cline{1-4}
188.6&3.21&$3.135\pm0.102$&0.85&Working Group\cite{lepsld} \\ \hline\hline
\end{tabular}
\label{tabla1}
\end{table}

For LEP2 let us define the ratio $R_{\tau\bar{\tau}}$ as:
\begin{eqnarray}
R_{\tau\bar{\tau}}\equiv \frac{\sigma(e^+e^-\rightarrow \tau^+\tau^-)}{
\sigma(e^+e^-\rightarrow \tau^+\tau^-)_{SM}}&=&1+ F_1^\gamma(s)\,  \eg + 
F_2^\gamma(s)\, \eg^2+F_1^Z(s)\, \ez \nonumber \\
&&+F_2^Z(s)\,  \ez^2 +F^{\gamma Z}(s)\, \ez \eg~.
\label{eq:ratiocross2}
\end{eqnarray}

For the range of energies used by LEP2 experiments, the
coefficients $F_i(s)$, obtained from \eqs{eq:angdist}{eq:coef}, are given in 
table \ref{tabla2}. Direct comparison of \eq{eq:ratiocross2} with 
experimental data will provide bounds on anomalous couplings. We have
checked that, even though coefficients in table \ref{tabla2} are obtained
with no initial state radiation, its inclusion only changes 
the coefficients by about a 10\% and  
this does not affect significantly the obtained bounds.

\begin{table}
\caption{Coefficients of the anomalous contributions to $R_{\tau\bar{\tau}}$, for 
the different center of mass measured energies at LEP2.}
\begin{tabular}{|c||c|c|c|c|c|}
\hline
$\sqrt{s}$(GeV)&$F^\gamma_1$&$F^\gamma_2$&$F^Z_1$&$F^Z_2$&
$F^{\gamma Z}$\\ \hline
\hline
130&0.079&0.682&0.028&5.258&0.286\\ \hline
136&0.083&0.784&0.026&5.152&0.304\\ \hline
161&0.092&1.221&0.022&5.272&0.384\\ \hline
172&0.094&1.427&0.021&5.497&0.424\\ \hline
183&0.096&1.642&0.020&5.789&0.467\\ \hline
189&0.096&1.765&0.019&5.971&0.491\\ \hline\hline
\end{tabular}
\label{tabla2}
\end{table}

In order to see how well the new couplings
can be bound from LEP2 let us find the limits on $\eg$ obtained
by using only the data at $189$~GeV (All data are used independently in the global fit discussed in section
\ref{sec:limits}). The
experimental value for the ratio $R_{\tau\bar{\tau}}$ is:
\begin{equation}
\left.R_{\tau\bar{\tau}}\right|_{\rm exp}=0.978\pm0.032~,
\end{equation}
which must be compared with our theoretical prediction (assuming that $\ez$ is
well bounded from LEP1-SLD, as it is)
\begin{equation}
\left.R_{\tau\bar{\tau}}\right|_{\rm th}=1.00+1.765\eg^2+0.096\eg~.
\end{equation}
From the two previous equations it is easy to find the following $1\sigma$ 
bound:
\begin{equation}
-0.10< \eg < 0.05~,
\label{eq:ratiocross2limit}                 
\end{equation}
which is comparable to the one obtained in
the final global fit given in section \ref{sec:limits}, where 
all available data have been included.

\subsection{Tau lepton transverse polarization asymmetry\label{sec:pol}}
Although magnetic moments change chirality, total cross sections 
are chirality even observables. Thus, in the limit of massless taus,
magnetic moment contributions to cross sections come always squared.
In addition all kind of new chirality even physics will also contribute
to total cross sections. Therefore, observables that vanish
for massless taus are superior because they depend linearly on magnetic
moments and therefore they are more sensitive to them.
On the other hand
they will not get contributions from physics conserving chirality.
In that sense they are truly magnetic moment observables. 

At LEP1\cite{acc}, with the $\tau$ direction fully reconstructed
in the semi-leptonic decays
$ e^+ e^- \rightarrow
\tau^+\  \tau^-\rightarrow h^+_1 X\; h^-_2 \nu_\tau\,$, 
$h^+_1 \bar{\nu_\tau}\; h^-_2 X$ ($h_1,h_2=\pi,\rho$), 
it has been shown\cite{nos}  that  one can  
get relevant information about the anomalous weak magnetic moment  just by
measuring  the following azimuthal asymmetry of the 
${\tau}$-decay products:

\begin{equation}
A_{cc}^\mp=\frac{\sigma^\mp_{cc}(+)-\sigma^\mp_{cc}(-)}
{\sigma^\mp_{cc}(+)+\sigma^\mp_{cc}(-)}~,\label{eq:asym}
\end{equation}

\noindent where $\sigma^\mp_{cc}$ is defined in the following angular regions

\begin{eqnarray}
\hspace*{-.4cm}\sigma^\mp_{cc}(+)&=&\sigma\left(\cos\theta_{\tau^-}>0,
\cos\phi_{h^\mp}>0\right)
+\sigma\left(\cos\theta_{\tau^-}<0,\cos\phi_{h^\mp}<0\right)~,\\
\hspace*{-.4cm}\sigma^\mp_{cc}(-)&=&\sigma\left(\cos\theta_{\tau^-}>0,
\cos\phi_{h^\mp}<0\right)
+\sigma\left(\cos\theta_{\tau^-}<0,\cos\phi_{h^\mp}>0\right)~.
\end{eqnarray}

In the $\beta\rightarrow1$ limit, the expression that one finds 
for the proposed asymmetry is:
\begin{eqnarray}
&&\spa A_{cc}^\mp= \mp \alpha_h 
\frac{1}{2}\frac{a}{v^2+a^2}
\left[  
-v\, r_Z +\epsilon_Z 
\right]~,\label{acc}
\end{eqnarray}
where $\alpha_h=(m_\tau^2-2m_h^2)/((m_\tau^2+2m_h^2)$, is the polarization
analyzer for each hadron channel ($h=\pi,\rho$), $a\equiv a_e=a_\tau$,
and $v\equiv v_e=v_\tau$. The formula shows that
$A_{cc}^\mp$,  being sensitive to the transverse polarization
of the $\tau$ lepton, selects the leading contribution in
the  anomalous weak magnetic coupling $\epsilon_Z$ of the tau.
In addition, the SM contribution to the observable is doubly suppressed
with respect to the non-standard one: by the
fermion vector coupling $v$ ($\equiv -\frac{1}{2}+2s_W^2$) and by the $r_Z$ (
$\displaystyle \equiv \frac{m_\tau}{m_Z})$ factor.

Within $1\sigma$, the
LEP1 \cite{acc} measurement of this asymmetry and the SLD\cite{sld}
determination of the transverse tau polarization, 
translate in the following values for the $\epsilon_Z$ coupling
\begin{equation}
\epsilon_Z= \left\{\begin{array}{l}
(0.0\pm1.7\pm2.4)\times 10^{-2}\;(\mathrm{LEP1})~,\cr
(0.28\pm1.07\pm0.81)\times 10^{-2}\; (\mathrm{SLD})
\end{array}\right.
\end{equation}

Combining these results one gets the bound:

\begin{equation}
\epsilon_Z= 0.002\pm 0.012~. \label{eq:ass1}
\end{equation}

Note that even though the transverse tau polarization has been 
measured at LEP1-SLD
with a precision one order of magnitude worse than the universality test
$R_{\tau \mu}$  (2-4\% typically for the asymmetry, and 0.5\% for the
tau-muon cross section ratio), the obtained bound \eq{eq:ass1} is
as good as the  one coming from universality \eq{eq:ratiocross1limit}.
This is  so because the asymmetry depends linearly on the couplings.
All the other observables depend mainly quadratically on the $\epsilon$'s,
therefore, if the new-physics  contributions to magnetic moments are
ever found to be different from zero, the asymmetry will be the only observable
able to disentangle the sign of the couplings.
In addition, as commented before, the
asymmetry  $A_{cc}^\mp$ is also qualitatively a better observable 
since it is independent on any physics that does not break chirality.

At present we do not know any similar measurement at LEP2. However we think
that this measurement will be very interesting since it will allow us to 
disentangle cleanly the $\gamma$ components from the $Z$ components of the 
magnetic moments.

\section{Lepton universality in $W \rightarrow \tau \nu_\tau$\label{sec:wdec}} 

 From \eq{eq:leff_fin} we see that the same couplings that give rise to
electromagnetic and $Z$-boson magnetic moments, also contribute  to the
couplings of the $W$ gauge bosons to tau leptons. The couplings appear in a
different combination than that in the photon or $Z$ couplings, so their
study gives us an additional independent information on magnetic moment
couplings. As was noticed in Ref. \cite{rizzo} the best place to look for
effects of the $\epsilon_W$ coupling is in the $W$ decay widths.

Using our effective Lagrangian we can easily compute the ratio of the decay
width of the W-gauge boson in tau-leptons (with magnetic moments) to
the decay width of the W to electrons (without magnetic moments).
\begin{eqnarray}
R^W_{\tau\e} &\equiv& \frac{\Gamma(W\rightarrow \tau\nu)}
{\Gamma(W\rightarrow e\nu)}\nonumber\\
&=&(1-r_W^2)^3\left[ 1+\frac{r_W^2}{2}+
3\sqrt{2}\; r_W\ c_W\, \epsilon_W 
+(1+2\, r_W^2)\epsilon_W^2\right]~,
\label{eq:ratiow}
\end{eqnarray}
where $r_W = m_\tau/m_W$, and $\epsilon_W$ can be rewritten 
in terms of $\epsilon_\gamma$ and $\epsilon_Z$ as in \eq{eq:epsilonw}. 
Similar expression to \eq{eq:ratiow} was given in Ref.~\cite{rizzo} but we 
found and extra global $(1-r_W^2)$ missing, and a factor
$3$ instead of a $\left(-\frac{3}{2}\right)$ factor (in our notation) in the term 
linear in the coupling.
Note that $R^W_{\tau\e}$, like the cross
sections studied in section \ref{sec:crossec}, is a chirality even observable.
Therefore, in the limit of massless taus, the only contribution from
magnetic moments comes squared.

The decay of the $W$ into leptons has been measured to a rather
good precision at LEP2, UA1, UA2, CDF and D0. There, results are presented 
in the form  of universality tests on the couplings (for a review on tau
lepton universality tests see Ref.~\cite{pich}) 
\begin{eqnarray}
\frac{g_\tau}{g_e} =&0.987 \pm 0.025  &\
\mathrm{(Colliders)}  \ \ \cite{tauw1}~,\label{eq:wuniv1}\\
\frac{g_\tau}{g_e} =&1.010 \pm 0.019  &\  
\mathrm{(LEP2)}   \ \ \cite{lepsld}~.
\label{eq:wuniv}
\end{eqnarray}
We combine these results and rewrite them as a measurement on the
ratio $R^W_{\tau\e}$ defined above
\begin{equation}
R^W_{\tau\e} = 1.002 \pm 0.030~,
\label{eq:ratiowexp}
\end{equation}
where we have assumed that the small effect (0.12\%) of the tau 
lepton mass has been subtracted to obtain the results shown in
\eq{eq:wuniv1} to \eq{eq:wuniv}.

Then, from \eq{eq:ratiowexp} and \eq{eq:ratiow},  
we obtain the following limit on the $W$-boson magnetic moments.
\begin{equation}
-0.23\leq \epsilon_W\leq 0.15~.
\label{eq:wlimit}
\end{equation}

\section{Combined limits on electromagnetic and weak magnetic moments of 
the tau lepton\label{sec:limits}}

We have performed a global fit, as a function of the two independent
couplings $\epsilon_\gamma$ and $\epsilon_Z$, to the  following studied 
observables : 
\begin{itemize}
\item ratio of cross sections 
$R_{\tau\mu}=\frac{\sigma(e^+e^-\rightarrow \tau^+\tau^-)}
                 {\sigma(e^+e^-\rightarrow
\mu^+\mu^-)}$ (\eq{eq:ratiocross}, Lepton Universality), at LEP1 and SLD (\eq{eq:ratiocross1}); 
\item  the ratio of cross sections
$R_{\tau\bar{\tau}}\equiv \frac{\sigma(e^+e^-\rightarrow \tau^+\tau^-)}{
\sigma(e^+e^-\rightarrow \tau^+\tau^-)_{SM}}$ (\eq{eq:ratiocross2}),
for the two highest energies measured at LEP2 (table \ref{tabla1});
\item the transverse tau polarization and the tau polarization asymmetry $A_{cc}^\mp=\frac{\sigma^\mp_{cc}(+)-\sigma^\mp_{cc}(-)}
{\sigma^\mp_{cc}(+)+\sigma^\mp_{cc}(-)}$ 
(\eq{eq:asym}) measured at SLD and LEP1 
(\eq{eq:ass1});
\item and the ratios of decay widths of W-gauge bosons
$R^W_{\tau\e} \equiv \frac{\Gamma(W\rightarrow \tau\nu)}
{\Gamma(W\rightarrow e\nu)}$
(\eq{eq:ratiow}) measured
at LEP2 and $p\bar{p}$ colliders (\eq{eq:ratiowexp}). 
\end{itemize}

\begin{figure}[hbtp]
\begin{center}
\epsfig{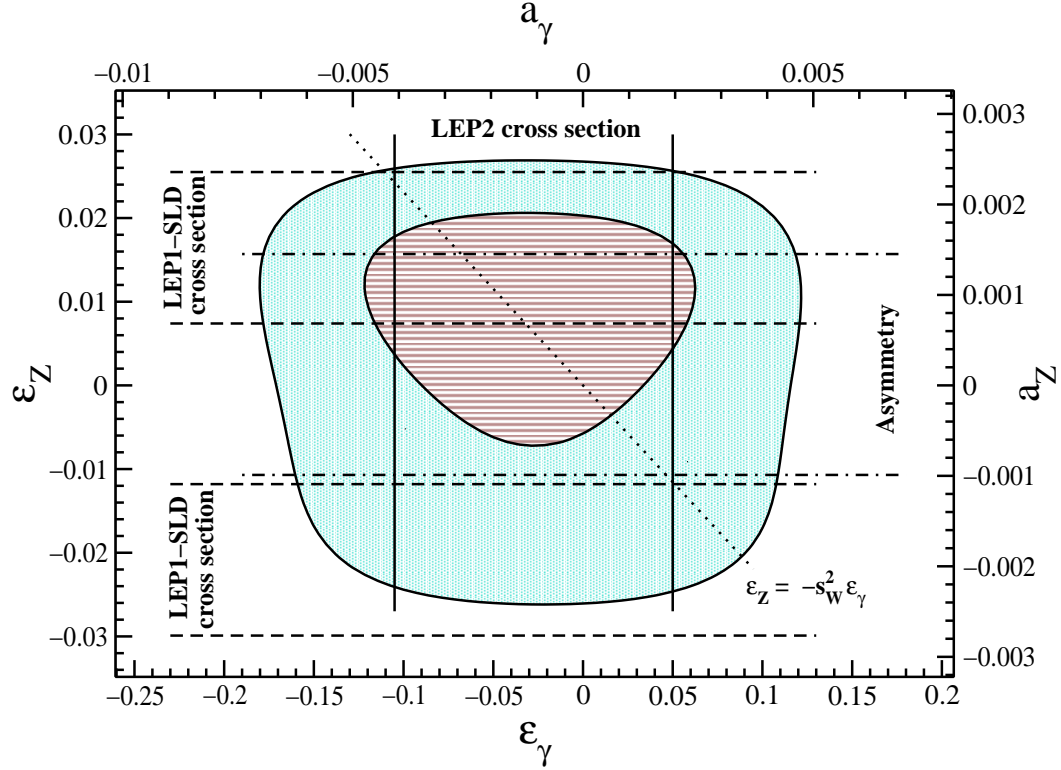}
\end{center}
\caption{Global fit including all constraints discussed in the paper.
95\% CL and 68\% CL contours are shown. The bands between straight lines 
show the allowed regions (1$\sigma$) coming from the different experiments: 
solid (LEP2-189 GeV), dashed (LEP1-SLD cross section), dot-dashed (asymmetry).
We also have plotted the line
$\epsilon_Z = - s^2_W \epsilon_\gamma$ (dotted line). 
This relationship appears when
only the operator ${\cal O}_B$ contributes.
\label{fig:gfit}}
\end{figure}

In fig.~\ref{fig:gfit} we present, in the plane $\eg$--$\ez$ (or
$a_\gamma$--$a_Z$)
the allowed region of parameters at 1$\sigma$ and 2$\sigma$.
For comparison we also present (at 1$\sigma$) the relevant limits set independently
by the different observables, as discussed in the text.
By projecting onto the axes one can read off the 1$\sigma$ and 2$\sigma$
limits on the different couplings
\begin{eqnarray}
&&(1\sigma)\rightarrow \left\{
\begin{array}{ll}
 -0.12< \epsilon_\gamma <0.06~, \\ 
 -0.0072< \epsilon_Z <0.021~,  
\end{array}
\right.\\
&&(2\sigma)\rightarrow \left\{
\begin{array}{ll}
 -0.18< \epsilon_\gamma <0.12~, \\ 
 -0.026< \epsilon_Z <0.027~.
\end{array}
\right.
\end{eqnarray}
At 1$\sigma$ the allowed region is far from elliptic. This is because the
dominant quadratic dependence of the observables on the parameters $\eg$ and
$\ez$ gives a tendency to provide two symmetric zones around two
different minima. This is especially true for  the bounds coming from
universality tests at the $Z$-peak. In this situation 
the interpretation of contours as
68\% CL contours is not clear. Probably one can combine the two minima
and obtain more stringent bounds. However in order to be conservative we
just quote the maximum allowed region. In any case, 2$\sigma$ contours do
not show this effect and we think they can be used reliably to obtain
95\% CL limits.

Using the relationship among $\epsilon_\gamma$, $\epsilon_Z$,
$\alpha_B$ and $\alpha_W$ at a given value of the scale of new physics,
one can easily obtain bounds on $\alpha_B$ and $\alpha_W$. Alternatively,
by assuming that $\alpha_B/4\pi$ or $\alpha_W/4\pi$ are order unity one can obtain
bounds on the scale of new physics $\Lambda$ ($\Lambda > 9$~TeV).

Finally the limits on $\epsilon_\gamma$ and $\epsilon_Z$ can be immediately
translated into limits on the non-standard contributions to 
the anomalous electromagnetic and weak
magnetic moments $a_\gamma$, $a_Z$ just by using \eq{eq:moments}. Thus we 
have:
\begin{eqnarray}
&&(1\sigma)\rightarrow \left\{
\begin{array}{ll}
 -0.005< a_\gamma <0.002~, \\ 
 -0.0007 < a_Z < 0.0019~,  
\end{array}
\right.\\
&&(2\sigma)\rightarrow \left\{
\begin{array}{ll}
 -0.007 < a_\gamma < 0.005~, \\ 
 -0.0024 < a_Z <0.0025~.  
\end{array}
\right.\label{eq:final2sigma}
\end{eqnarray}
For $a_\gamma$ these limits are only about one order of magnitude larger
than the standard model contribution, $a_\gamma^{\mathrm SM} \sim 0.00117$.

\section{Discussion and conclusions \label{sec:conc}}

The above bounds are completely model independent and no assumption has 
been made on the relative size of couplings $\alpha_B$ and $\alpha_W$ in the
effective Lagrangian (\ref{eq:leff}). For the sake of comparison with
published data \cite{masso} we present now the limits that can
be found by considering separately  only operator ${\cal O}_B$ or only
operator ${\cal O}_W$ in the Lagrangian (\ref{eq:leff}). Consider that only
${\cal O}_B$ is present, as in Ref.~\cite{masso}, is equivalent 
to impose the relation $\ez=-s_W^2\eg$. Thus, from fig.~\ref{fig:gfit}, it is
straightforward to obtain that the bounds on the anomalous magnetic 
moment (at $2\sigma$) are reduced to $-0.004 < a_\gamma < 0.003$, while little 
change is found on the weak-magnetic moment $-0.0019 < a_Z < 0.0024$.

\begin{figure}[hbtp]
\begin{center}
\epsfig{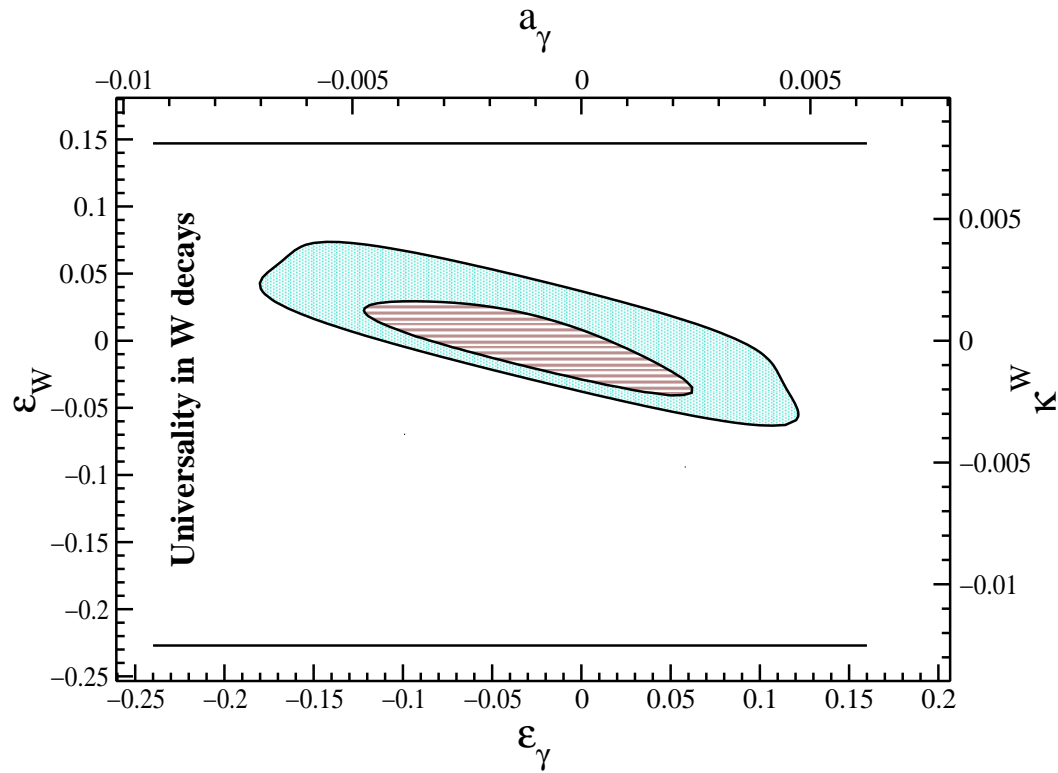}
\end{center}
\caption{The global fit of fig.~\ref{fig:gfit} is now plotted in the
plane $\eg$ and $\epsilon_W$ to show the combined bounds on $\epsilon_W$.
As in fig.~\ref{fig:gfit} 95\% CL and 68\% CL contours are shown. 
For comparison we also
draw as straight lines the direct 1$\sigma$ bounds obtained from
universality tests in $W$ decays.
\label{fig:gfitw}}
\end{figure}

Universality tests in $W$ decays do not provide any interesting constraint
on the couplings $\ez$ and $\eg$. However, because of the relationship
(\ref{eq:epsilonw}), the LEP1-SLD and LEP2 constraints on $\ez$ and $\eg$ can
be translated into constraints on $\epsilon_W$ (or
$\kappa^W$ defined in \eq{eq:moments}),
the weak magnetic moment couplings of the $W$-gauge-boson to taus and
neutrinos. In fig.~\ref{fig:gfitw} we present,
in the plane $\eg$--$\epsilon_W$ (or $a_\gamma$--$\kappa^W$),
the $1\sigma$ and $2\sigma$ regions obtained from our global fit to all data. Clearly
LEP1-SLD limits on $\ez$ coming from $Z$-decays and the asymmetry and LEP2
limits on $\eg$ constrain $\epsilon_W$ very strongly. For comparison we
also plot, as straight lines, the 1$\sigma$ limits we have obtained from the
universality tests in $W$ decays. 

~From the figure \ref{fig:gfitw}, one immediately obtains the 95\% CL limits
\begin{equation}
-0.06 < \epsilon_W < 0.07, \ \ \ {\mathrm or\ equivalently}\ \ \ 
-0.003 < \kappa^W < 0.004~.\label{eq:kweak}
\end{equation}

Bounds on the anomalous electromagnetic moment
of the $\tau$ can also be obtained from the radiative 
decay $Z\rightarrow \tau^+\ \tau^- \
\gamma$ at LEP1 \cite{mendez,riemann}. 
There, only the anomalous coupling $a_\gamma$ is
taken into account, while the contributions coming
from the tau $Z$-magnetic coupling $a_Z$ are neglected.
Using this approach,  with the
inclusion of linear terms in $a_\gamma$ in the cross section 
\cite{riemann},  and taking into account that the $\tau$ which emits 
the photon is off-shell, the analysis of the L3 
\cite{L3} and OPAL \cite{opal} collaborations lead to the 
limits (at $95\%$ CL):
\begin{eqnarray}
&&-0.056<a_\gamma<0.044\ \ \ \mathrm{[L3]}~,\\
&&-0.068<a_\gamma<0.065\ \ \ \mathrm{[OPAL]}~.
\end{eqnarray}

As can be seen from \eq{eq:final2sigma} our result, coming mainly from
LEP2, is about one order of magnitude better than the ones obtained 
from the radiative $Z$-decay.
In both cases some
of the particles in the vertex are off-shell. The interpretation
of off-shell form factors is problematic since they can hardly be isolated
from other contributions and gauge invariance can be a problem. In
the effective Lagrangian approach all those problems are solved because
form factors are directly related to couplings in the effective Lagrangian,
which is gauge invariant, and as discussed in section~\ref{sec:el},
the difference $F^{\mathrm new}(q^2)-F^{\mathrm new}(0)$ only comes from
higher dimension operators whose effect is suppressed by $q^2/\Lambda^2$.

Concluding, we have shown that the use of all available data at the highest
available energies (LEP1, SLD, LEP2, D0, CDF) allowed us to strongly 
constrain all the magnetic moments (weak and electromagnetic) of the tau lepton
without making any assumption about naturalness or fine tuning. The obtained
bounds (\eq{eq:final2sigma} and \eq{eq:kweak}), to our knowledge,  are 
the best bounds that one can find in published data.

\begin{ack}
This work has been supported by CICYT under the 
Grant AEN-99-0692, by DGESIC under 
the Grant PB97-1261, by the DGEUI of 
the Generalitat Valenciana under the Grant GV98-01-80,
by Agencia Espa\~nola de Cooperaci\'on Internacional and by CSIC-Uruguay.
\end{ack}


\begin{thebibliography}{20}
\bibitem{masso} R.~Escribano and E.~Mass\'o, \pl{B301}{93}{419} and \pl{B395}{97}{369}.
\bibitem{pdg} C.~Caso \etal, \epj{C3}{1998}{1}.
\bibitem{masso2} R.~Escribano and E.~Mass\'o, \np{B429}{94}{19}
\bibitem{buch}W.~Buchmuller and D.~Wyler, \np{B268}{86}{621};
C.N.~Leung, S.T.~Love and S.~Rao, \zp{C31}{86}{433};
M. Bilenky and A. Santamaria, \np{B420}{94}{47}.
\bibitem{rizzo} T.G. Rizzo, \pr{D56}{97}{3074}.
\bibitem{lowbounds} M.T. Dova, J. Swain and L. Taylor, \pr{D58}{98}{015005};
M.T. Dova, P. Lacentre, J. Swain and L. Taylor, \jhep{9907}{99}{026}.
\bibitem{nos} J. Bernab\'eu, G. Gonz\'alez-Sprinberg and J. Vidal, 
\pl{B326}{94}{168};
J. Bernab\'eu, G. Gonz\'alez-Sprinberg, M. Tung and J. Vidal, 
\np{B436}{95}{474};
J. Vidal, J. Bernab\'eu and G. Gonz\'alez-Sprinberg, \npp{B76}{99}{221}.
\bibitem{acc} L3 Collaboration, M. Acciarri \etal, \pl{B426}{98}{207}.
\bibitem{sld}SLD Collaboration, Kenji Abe, \etal, SLAC-PUB-8163.
\bibitem{lepsld} LEP Collaborations ALEPH, DELPHI, L3, OPAL, the 
LEP Electroweak  Groups and the SLD Heavy Flavour and  Electroweak Groups, 
CERN-EP/2000-016. 
\bibitem{bernabeu} A.A. Akhundov, J. Bernab\'eu, D. G\'omez-Dumm and A. Santamaria, 
\np{B563}{99}{82}.
\bibitem{all} OPAL Collaboration, G. Abbiendi \etal, \epj{C6}{1999}{1},
CERN-EP/99-097 (hep-ex/9908008); ALEPH Collaboration, R. Barate \etal, 
\epj{C12}{2000}{183};  L3 Collaboration, M. Acciarri \etal, CERN-EP/99-181 
(hep-ex/0002034).
\bibitem{pich} A. Pich, hep-ph/9912294.
\bibitem{tauw1} S. Protopopescu, in the {\it Proceedings of the Fifth
International Workshop on Tau Lepton Physics}. Santander, Spain, September
1998, \npp{B76}{99}{91}; J. Ellison, Proc. EPS-HEP99 Conference, Tampere,
July 1999, hep-ex/9910037.
\bibitem{mendez}J.A. Grifols and A. M\'endez, \pl{B255}{91}{611}.
\bibitem{riemann}J. Biebel and T. Riemann, \zp{C76}{97}{53}; S.S. Gau \etal,
\np{B523}{98}{439}
\bibitem{L3} M. Acciarri \etal, \pl{B434}{98}{169}.
\bibitem{opal} K. Ackerstaff \etal, \pl{B431}{98}{188}.
\end{thebibliography}
\end{document}